\documentclass[aps,prd,superscriptaddress,showpacs,preprintnumbers]{revtex4}
\usepackage{graphicx}

\begin{document}

\newcommand{\dlt}{\bigtriangleup}
\newcommand{\beq}{\begin{equation}}
\newcommand{\eeq}[1]{\label{#1} \end{equation}}
\newcommand{\insertplot}[1]{\centerline{\psfig{figure={#1},width=14.5cm}}}

\parskip=0.3cm


\title{EXCLUSIVE $J/\Psi$ ELECTROPRODUCTION IN A DUAL MODEL}

\author{R. Fiore}
\affiliation{Dipartimento di Fisica, Universit\`{a}
della Calabria and \\ Instituto Nazionale di Fisica
Nucleare, Gruppo collegato di Cosenza
 \\
I-87036 Arcavacata di Rende, Cosenza, Italy}

\author{L.L. Jenkovszky}
\affiliation{Bogolyubov Institute for Theoretical Physics (BITP),
Ukrainian National Academy of Sciences \\14-b, Metrolohichna str.,
Kiev, 03680, Ukraine}

\author{V.K.~Magas}
\affiliation{Departament d'Estructura i Constituents de la
Mat\'eria,\\ Universitat de Barcelona, Diagonal 647,\\
08028 Barcelona, Spain}

\author{S.~Melis}
\affiliation{Di.S.T.A., Universit\`a del Piemonte Orientale
``A. Avogadro'', 15100
Alessandria, Italy, \\
          INFN, Sezione di Torino, Via P. Giuria 1, I-10125 Torino, Italy}

\author{A.~Prokudin}
\affiliation{Dipartimento di Fisica Teorica, Universit\`a di Torino and \\
          INFN, Sezione di Torino, Via P. Giuria 1, I-10125 Torino, Italy}
\affiliation{Jefferson Laboratory, 12000 Jefferson Avenue, Newport News, VA 23606}

\begin{abstract}
Exclusive $J/\Psi$ electroproduction is studied in the framework
of the analytic $S-$matrix theory. The differential and integrated
elastic cross sections are calculated using the Modified Dual
Amplitude with Mandelstam Analyticity (M-DAMA) model. The model is
applied to the description of the available experimantal data and
proves to be valid in a wide region of the kinematical variables
$s$, $t$ and $Q^2$. Our amplitude can be used also as a universal
background  parametrization
for the extraction of tiny resonance signals.
\end{abstract}


\maketitle

\section{Introduction} \label{s1}
High-energy exclusive $J/\Psi$ electroproduction was intensively
studied in recent years. The theoretical tools include the quark
model, perturbative QCD and Regge pole models (for relevant
reviews see Ref. \cite{review}). The main source of the data at
high energies is the HERA collider at DESY~\cite{breitweg99,adloff99,chekanov02,chekanov04,aktas06}, while
at low energies experimental studies at JLab are promising, see
for instance the recent paper, Ref.~\cite{JLab}.

In Ref.~\cite{our_paper} a model based on vector meson dominance
(VMD) and a dual amplitude with Mandelstam analyticity (DAMA)
\cite{DAMA} was proposed to describe $J/\Psi$ photoproduction; it
resulted in a very good description of the data in the region of
$s$ from the threshold to high energy, where pure Pomeron exchange
dominates. $J/\Psi$ photoproduction is an ideal tool to study the
Pomeron exchange.

In the framework of the VMD model \cite{VMD}, the photoproduction
scattering amplitude is proportional to the sum of the relevant
hadronic amplitudes
 \beq
 A(\gamma\, p\rightarrow V\, p)=\sum_V{e\over{f_V}}A(V\, p\rightarrow V\, p),
 \eeq{VMD}
where $V=\rho, \omega, \phi, J/\Psi, ...$, $f_V$ is a decay constant of the corresponding vector meson, and  $e$ is the vector meson-photon coupling constant.
Comparison with the experimental data
\cite{VMD_exp} confirms the validity of VMD in photoproduction
reactions. VMD in electroproduction becomes more complicated and
less obvious because of the asymmetry between the photon
virtuality and the produced vector meson mass. A ``skewed
symmetric" VMD model for vector meson electroproduction was
developed and applied in Ref. \cite{Fraas, Schild, Goeke, Ryskin},
while explicit Regge-pole models  were constructed recently in
Refs. \cite{DL, Mullerfit, Szczepaniak, Capua}.

VMD reduces $J/\Psi$ photo- (electro-)production $\gamma(\gamma^*)
p\rightarrow J/\Psi({J/\Psi}^\star) \, p$ to a purely hadronic
process $J/\Psi\, p\rightarrow J/\Psi\, p$ \cite{VMD}. Due to the
Okubo-Zweig-Iizuka rule \cite{OZI} and the two-component duality
picture \cite{H_R},  the high-energy Regge behaviour of this
scattering amplitude is completely determined by a Pomeron or,
equivalently, by a reggeized two-gluon ladder exchange, while its
low-energy behavior is that of a smooth background due to the
exotic quantum numbers of the direct channel. To a lesser extent,
this is true also for $\phi-p$ scattering, however in the latter
case ordinary meson exchange is present due to $\omega-\phi$
mixing. Heavier vector mesons are as good as $J/\Psi-p$, but
relevant data are less abundant. So, we find $J/\Psi-p$ scattering
to be an ideal testing field for diffraction, where it can be
studied uncontaminated by secondary trajectories \cite{our_paper}.
A detailed introduction to the OZI rule and the two-component
duality, both relevant to the present discussion, can be found in
Ref.~\cite{our_paper}.

In this paper we study electroproduction of $J/\Psi$ meson in the
process $\gamma^*(Q^2) p \rightarrow J/\Psi p$ that gives a unique
opportunity to study the properties of the Pomeron at different
scales $Q^2$. We extend our model to processes with virtual
particles based on  a modified version of DAMA, M-DAMA.
Modification of the amplitude is done in such a way that all
properties and symmetries of on-mass shell amplitude are secured.

The paper is organized as follows: in Section~\ref{s2} we introduce
the model, while in Section~\ref{s3} is dedicated to $J/\Psi$
electroproduction and description of the available experimental
data. Our conclusions are drawn in Section~\ref{s4}.

\section{THE MODEL} \label{s2}

With the advent of the HERA experiments, the Regge-pole model was
applied to off-mass shell processes by making some of its
parameters, e.g. the Pomeron intercept, $Q^2-$dependent. When going
back to low energies, two problems should be solved. One is the
construction of a relevant dual amplitude, interpolating between
low- and high energies and the second is its off-mass shell
continuation.

To solve the first problem in Ref. \cite{our_paper} we applied the
Dual Model with Mandelstam Analiticity (DAMA) to $J/\psi$
photoproduction. The parameters of the model were fitted to the
experimental data on the differential cross section, ${d\sigma /
dt}$. With these parameters we calculated also the total $J/\Psi$
photoproductioncross section, as a model prediction, and obtained
good agreement with the experimental data.

As to the second problem, we extend our analysis to off-shell
$J/\Psi p$ scattering (electroproduction). A Modified-DAMA
(M-DAMA) formalism including $Q^2-$dependence was proposed in
Ref.~\cite{MDAMA}. Our strategy is to keep fixed the parameters
obtained in the on-shell case, Ref. \cite{our_paper}, and to fit
the data on $J/\Psi$ electroproduction in the generalized model,
varying only the new parameters connected with the off-shell
continuation of the model \cite{MDAMA, jpsi2}.

Let us briefly review the main properties of our dual model.

\subsection{Kinematics}
\label{s_s_kin}

The photoproduction scattering amplitude with the use of VMD is
\cite{our_paper} \beq A_{\gamma p \rightarrow J/\Psi p}(s,t) = c_V
A_{J/\Psi p \rightarrow J/\Psi p}(s,t)  \,, \eeq{a0} where $c_V$
is the $\gamma-J/\Psi$ coupling, and we use the following
kinematic relation between Mandelstam variables \beq
 s+t+u=2 M_p^2 + 2 M_{J/\Psi}^2\,,
\eeq{kin0-r} corresponding to the VMD amplitude on the r.h.s. of
Eq. (\ref{a0}).

Similarly, for the case of electroproduction (i.e. photoproduction
by virtual photons) we write our generalized scattering amplitude
in the form
 \beq A_{\gamma^* p \rightarrow
J/\Psi p}(s,t,Q^2) = c_V A_{J/\Psi^* p \rightarrow J/\Psi
p}(s,t,Q^2)  \,, \eeq{a1} $Q^2$ is the photon virtuality.
The kinematic relation is also modified to meet the r.h.s. of Eq.
(\ref{a1}): \beq
 s+t+u=2 M_p^2 + 2 M_{J/\Psi}^2-Q^2\,.
\eeq{kin1-r} In the photoproduction limit ($Q^2\rightarrow 0$)
Eqs.~(\ref{a1},\ref{kin1-r}) reproduce
Eqs.~(\ref{a0},\ref{kin0-r}).

\subsection{Dual amplitude}
\label{s_dual}

The dual model with Mandelstam analyticity (DAMA) appeared as a
generalization of narrow-resonance (e.g. Veneziano) dual models,
intended to overcome the manifestly non-unitarity of the latter
\cite{DAMA}. Contrary to narrow-resonance dual models, DAMA requires  non-linear, complex
trajectories. The dual properties of DAMA were studied
in Ref. \cite{Jenk}.

The DAMA amplitude \cite{DAMA} is given
by:
\beq D(s,t)=c \int_0^1 {dz \biggl({z \over g}
\biggr)^{-\alpha(s')-1} \biggl({1-z \over
g}\biggr)^{-\alpha_t(t'')}}\,,
\eeq{dama_eq}
where $\alpha(s)$ and
$\alpha(t)$ are Regge trajectories in the $s$ and $t$ channel
correspondingly; $x'=x(1-z), \ \ x''=xz$ $(x=s,t,u)$; $g$ and $c$ are
parameters, $g>1$, $c>0$.

For $s\rightarrow\infty$ and fixed $t$ DAMA is
Regge-behaved
\beq
D(s,t)\sim s^{\alpha_t(t)-1}\,.
\eeq{dama_regge}
The pole structure of DAMA is similar to that of the Veneziano
model except that multiple poles appear on daughter levels
\cite{DAMA}:
\beq D(s,t)=\sum_{n=0}^{\infty}
g^{n+1}\sum_{l=0}^{n}\frac{[-s\alpha'(s)]^{l}C_{n-l}(t)}
{[n-\alpha(s)]^{l+1}}\,,
\eeq{series}
where $C_n(t)$ is the
residue, whose form is fixed by the $t$-channel Regge trajectory
(see \cite{DAMA}).
The pole term
in DAMA is a generalization of the Breit-Wigner formula,
comprising a whole sequence of resonances lying on a complex
trajectory $\alpha(s)$. Such a "reggeized" Breit-Wigner formula
has little practical use in the case of linear trajectories,
resulting in an infinite sequence of poles, but it becomes a
powerful tool if complex trajectories with a limited real part and
hence a restricted number of resonances are used.

A simple model of trajectories satisfying the threshold and
asymptotic constraints of DAMA is a sum of square roots \cite{DAMA}
$\alpha(s)\sim \sum_i\gamma_i\sqrt{s_i-s}$.
The number of thresholds included depends on the model; the simplest is with a single
threshold
\beq
\alpha(s)=\alpha(0)+\alpha_1(\sqrt{s_0}-\sqrt{s_0-s})\, .
\eeq{trajectory}
Imposing an upper bound on the real part of this
 trajectory,
 \beq
 Re\, \alpha(s)<0\, \quad \Rightarrow \quad \alpha(0)+\alpha_1\sqrt{s_0}<0\,,
 \eeq{Re_alpha}
 we get an
amplitude that does not produce resonances \cite{DAMA,superbroad,Jenk}.
The imaginary part of such a trajectory instead rises
indefinitely, contributing to the total cross section with a
smooth background.
This ansatz for the exotic trajectory was suggested in Ref.~\cite{our_paper}.

\subsection{Regge trajectories}

For the  $t$-channel Pomeron trajectory we use an expression with
two square roots thresholds: \beq
\alpha(t)=\alpha^P(t)=\alpha^P(0)+\alpha^P_1(\sqrt{t_1}-\sqrt{t_1-t})+
2\alpha^P_2(t_2-\sqrt{(t_2-t)t_2}) \eeq{Pomer} with a light
(lowest) threshold $t_1=4m_{\pi}^2$, while the value of the  heavy
one, $t_2$,  were fitted to the photoproduction data together with
the other parameters of the trajectory \cite{our_paper}
 (see Table \ref{fitpar}).

The direct-channel exotic trajectory is given by eq. (\ref{trajectory}),  with condition (\ref{Re_alpha}),
where the relevant threshold value is
$s_0=(m_{J/\Psi}+m_P)^2$. The other parameters of the trajectory - $\alpha(0)$ and $\alpha_1$ - were fitted to
the data in Ref. \cite{our_paper} (see Table \ref{fitpar}).

\subsection{Off mass shell Dual Model} \label{s_off}
To extend our model off-mass shell we need to construct the $Q^2$-dependent dual amplitude. To this aim we use
the so called Modified DAMA (M-DAMA) formalism developed in Ref.~\cite{MDAMA}.
The scattering amplitude is given by
\beq
D(s,t,Q^2)=c \int_0^1 {dz \biggl({z \over g}
\biggr)^{-\alpha(s')-\beta({Q^2}'')-1} \biggl({1-z \over
g}\biggr)^{-\alpha_t(t'')-\beta({Q^2}')}}\,,
\eeq{mdama_eq} where
 $\beta(Q^2)$ is a monotonically
decreasing dimensionless function of $Q^2$;
$x'=x(1-z), \ \ x''=xz$, where $x \equiv s, Q^2, t$.

It has been shown in Ref. \cite{MDAMA} that by choosing the $\beta$
function in the form
 \beq \beta(Q^2)=-{\alpha_t(0)\over \ln g}\ln
\left(\frac{Q^2+Q_0^2}{Q_0^2}\right)\,, \eeq{beta}
all asymptotics of the amplitude at large $s$ remain valid and $Q^2$ behaviour of the amplitude is in qualitative agreement with the experiment. Clearly at $Q^2 = 0$ we have
$\beta(0) = 0$ so that we reproduce the on-mass shell amplitude
studied in Ref.~\cite{our_paper}.

Now we can calculate the amplitude $D(s,t,Q^2)$ using the same numerical method as in Ref. \cite{our_paper}.
The replacement of DAMA by M-DAMA results only in one more
parameter, namely the characteristic virtuality scale $Q_0^2$.

\subsection{Scattering amplitude and cross sections}

From the amplitude $D(s,t,Q^2)$, Eq.~(\ref{mdama_eq}) we can
construct the full scattering amplitude. Following Refs.
\cite{our_paper} and \cite{MDAMA, jpsi2}, in order to secure ($s -
u$) symmetry, the total amplitude can be written as
\beq
A(s,t,u,Q^2)=(s-u)(D(s,t,Q^2)-D(u,t,Q^2))\,.
\eeq{A}
Although, in
the off-shell case, this symmetry is progressively violated as
$Q^2$ increases, we try this form as the first approximation
\cite{jpsi2}.

For the exotic Regge trajectory without resonances, like the one
given by eq. (\ref{trajectory}) with condition (\ref{Re_alpha}),
the scattering amplitude $A(s,t,u,Q^2)$,  eq. (\ref{A}), is given
by a convergent integral, and can be calculated for any $s$, $t$
and $Q^2$ without analytical continuation, needed otherwise, as
discussed in \cite{DAMA,MDAMA}.

\begin{table}[t]
\begin{center}
\begin{tabular}{|rl|rl|}
\hline
\multicolumn{4}{|c|}{{\bf ON-MASS-SHELL}  Ref.~\cite{our_paper}}\\
\hline
$\alpha^E(0)$ = & $-1.83$ & $\alpha^E_1(0)$ = & $0.01$ (GeV$^{-1}$) \\
$\alpha^P(0)$ = & $1.2313$ & $\alpha^P_1(0)$ = & $0.13498$ (GeV$^{-1}$)\\
$\alpha^P_2(0)$ = & $0.04$ (GeV$^{-2}$) & $t_2$ = & $36$ (GeV$^2$) \\
$g$ = & $13629$ & $c$ = &  $0.0025$ (GeV$^{-2}$)  \\
\hline
\multicolumn{4}{|c|} {$\chi^2/{d.o.f.}$ =  $0.83$}\\
\hline
\multicolumn{4}{|c|} {\bf OFF-MASS-SHELL}\\
\hline
\multicolumn{4}{|c|} {$Q_0^2$ = $3.464^2$ (GeV$^2$),  $\quad a$ =  $2.164$,  $\quad n$ =  $2.131$ }\\
\hline
\multicolumn{4}{|c|} {$\chi^2/{d.o.f.}$ =  $1.2$}\\
\hline
\end{tabular}
\end{center}
\vspace{-0.5cm}
\caption{Fitted values of the adjustable parameters
\label{fitpar}}
\end{table}

\section{$J/\Psi$ electroproduction}
\label{s3}

The transverse differential cross section is given by
\beq
{d\sigma_T\over{dt}}(s,t,Q^2)={1\over16\pi\lambda (s,
m_{J/\Psi}^2, m_P^2)}\vert A(s,t,u,Q^2)\vert ^2\,,
\eeq{dsigmadt}
where
$\lambda (x,y,z) = x^2 + y^2 + z^2 - 2xy - 2yz - 2xz$.

\begin{figure}{}
\includegraphics[width=.45\textwidth,bb= 10 140 540 660]{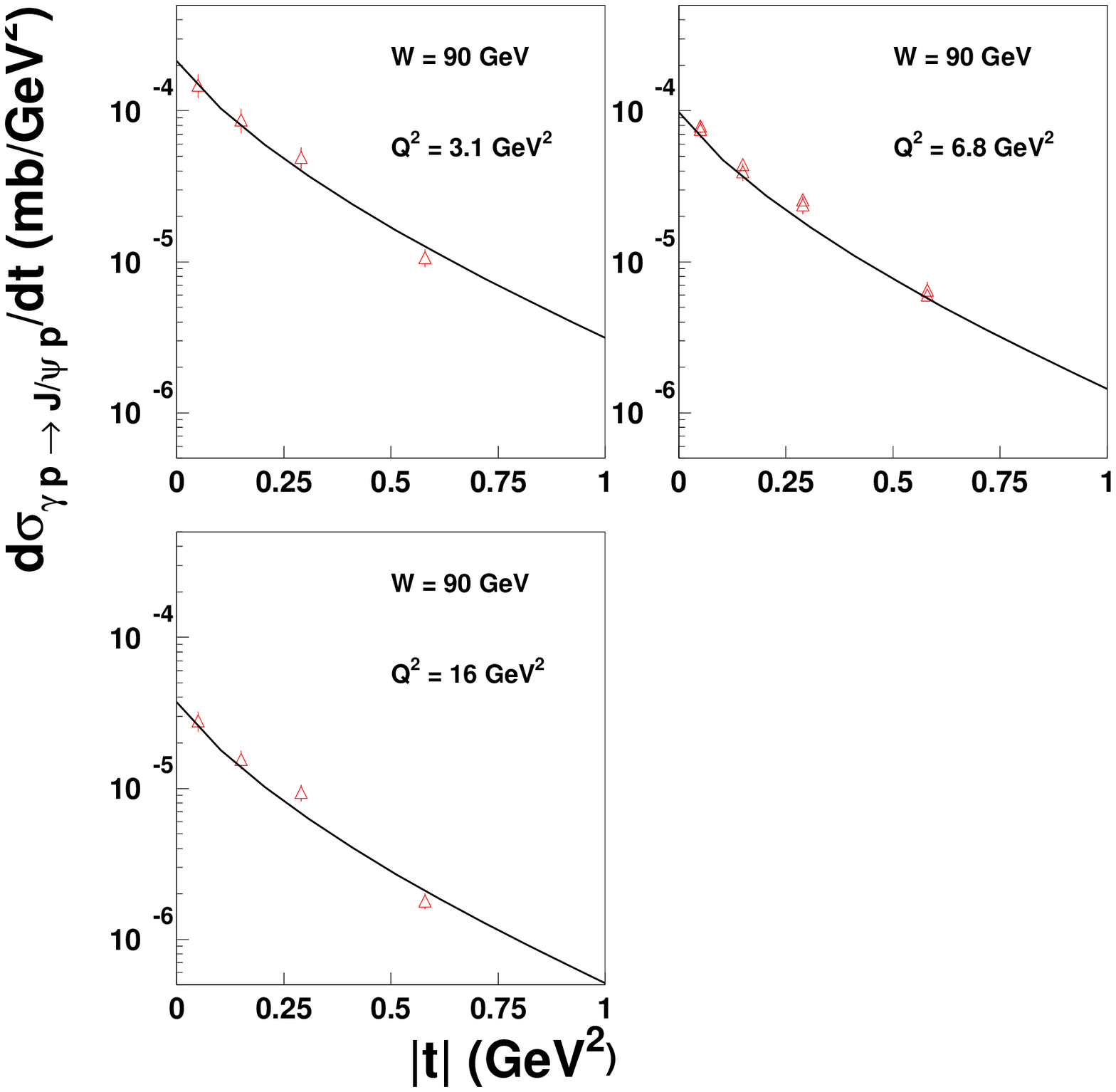}
\includegraphics[width=.45\textwidth,bb= 10 140 540 660]{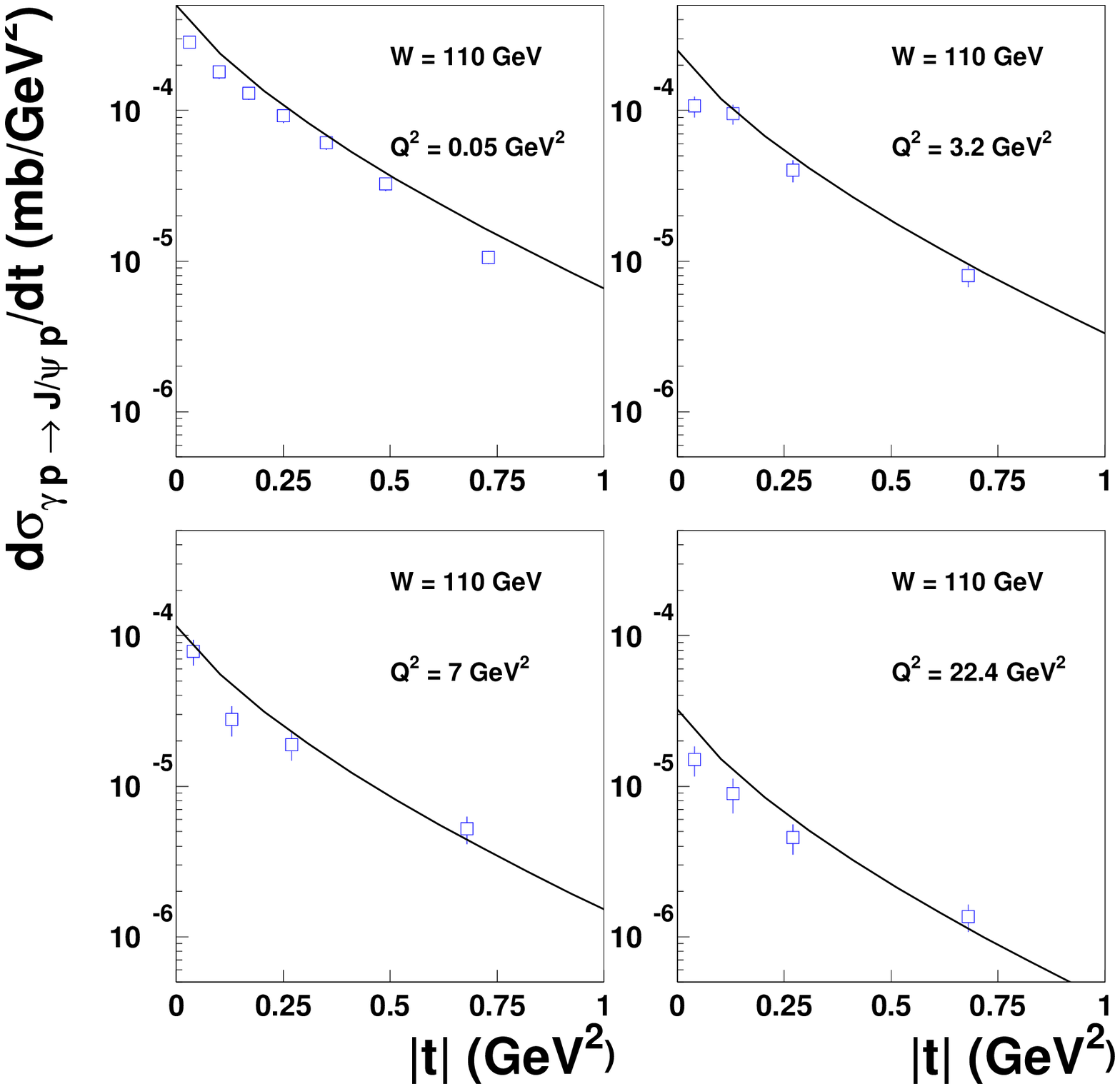}
\caption{$J/\Psi$ differential cross sections as a function of $t$  for different $Q^2$ for:\\
left panel - $W = 90$ GeV, experimental points are from Ref.  \cite{chekanov04};\\
right panel - $W = 110$ GeV, experimental points are from Ref.~\cite{aktas06}.  }
\label{fig1}
\end{figure}

The total transverse cross section reads
\beq
\sigma_T(s,Q^2)=\int_{-t_{max}=s/2}^{t_{thr.}\approx 0}dt\, {d\sigma_T\over{dt}}(s,t,Q^2)\,.
\eeq{sigma_T}

When calculating the total elastic cross section it is important
to take into account also the longitudinal component of the cross
section, which was negligible for the photoproduction case
\cite{our_paper}. Thus, the total elastic cross section is given
by the sum of longitudinal and transverse components: \beq
\sigma_{el}(s,Q^2)=(1+R(Q^2))\sigma_T(s,Q^2)\,, \eeq{sigma_el}
where $R=\sigma_L/\sigma_T$.

In Ref. \cite{martynov} the following expression for the
$R_{V}(Q^2)$, universal for all studied vector mesons $V \equiv
\rho_0, \phi, \omega, J/\Psi$, was proposed: \beq
R_{V}(Q^2)=\left(\frac{a\, m_{V}^2 + Q^2}{a\,
m_{V}^2}\right)^{n}-1\,, \eeq{Rq} where $a$ and $n$ are the
adjustable parameters. For our particular case, we choose
$R=R_{J/\Psi}(Q^2)$.


Thus, in order to describe $J/\Psi$ electroproduction data we have
three adjustable parameters: $Q_0^2$, $a$, and $n$. In the fitting procedure we include
76 data points on $\sigma_{el}$, 32 data points on $d\sigma/dt$ and 8 data points on $R=R_{J/\Psi}(Q^2)$ at $Q^2\ne 0$ from Refs~\cite{breitweg99,adloff99,chekanov02,chekanov04,aktas06}. The resulting values of fitted
parameters are: $Q_0^2=3.464^2$ (GeV$^2$), $a=2.164 $, $n=2.131$. All other
parameters are fixed to the values obtained in Ref.
\cite{our_paper}. The agreement is very good,
$\chi^2/d.o.f.=1.2$ for 116 data points, as seen in Table \ref{fitpar}.

\begin{figure}{}
 \includegraphics[width=.45\textwidth,bb= 10 140 540 660]{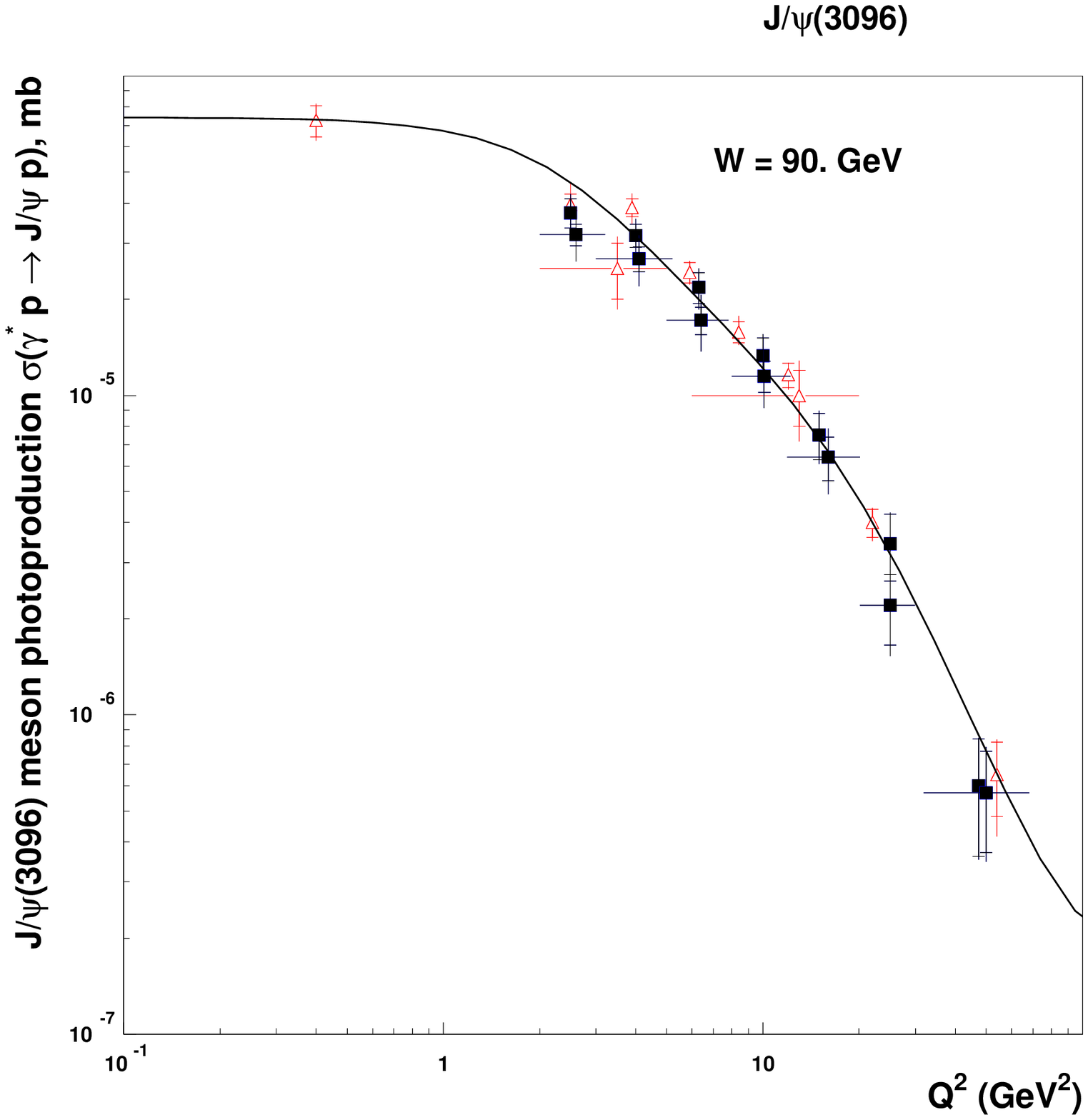}
 \includegraphics[width=.45\textwidth,bb= 10 140 540 660]{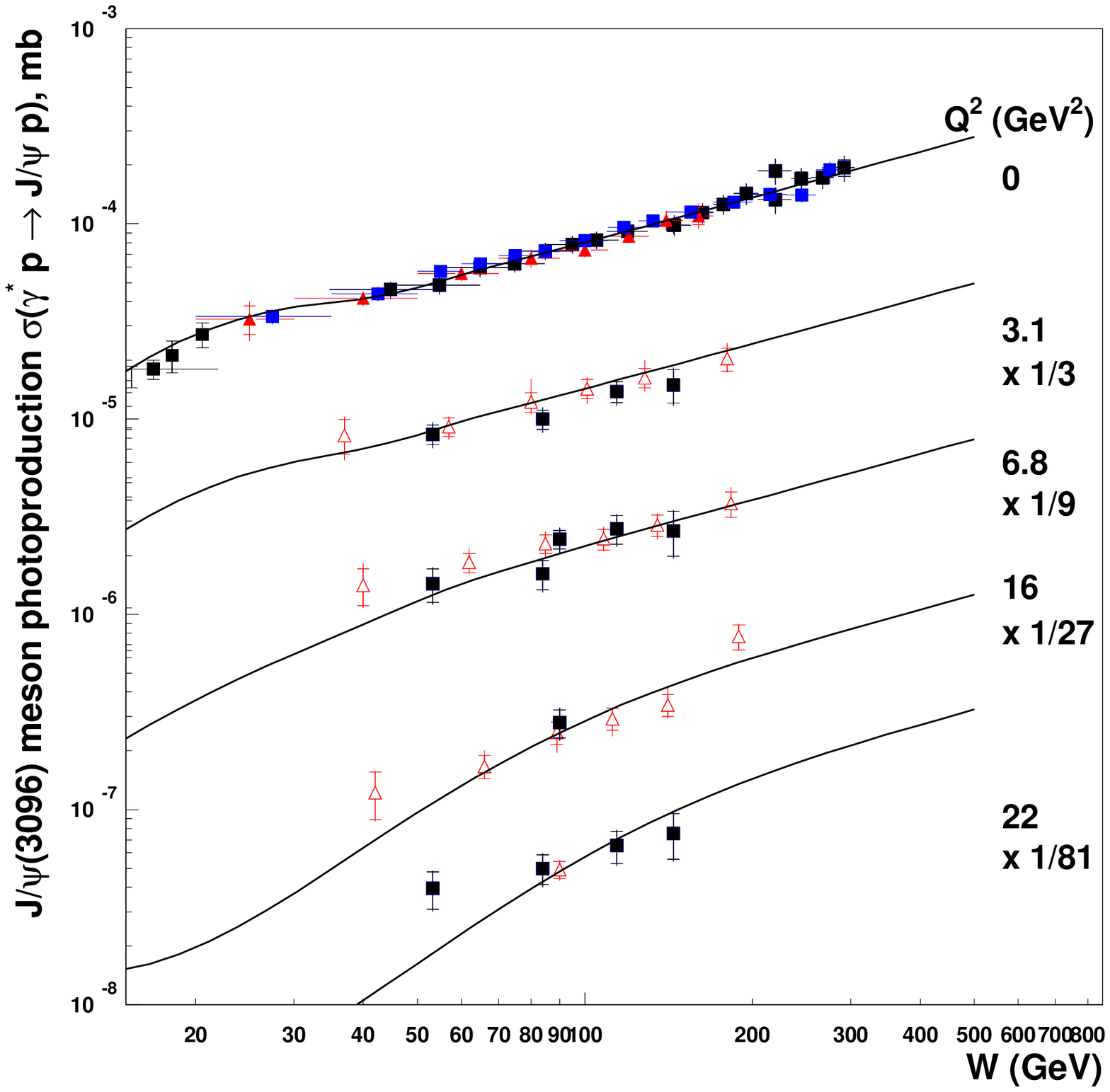}
\caption{ $J/\Psi$ elastic cross
sections  as a function of $Q^2$  for $W=90$ GeV  (left panel); and
as a function of energy  $W$ for different $Q^2$ (right panel).  Experimental points are from Refs. ~\cite{adloff99,breitweg99,chekanov02,chekanov04,aktas06}.}
\label{fig2}
\end{figure}

\begin{figure}{}
 \includegraphics[width=.45\textwidth,bb= 10 140 540 660]{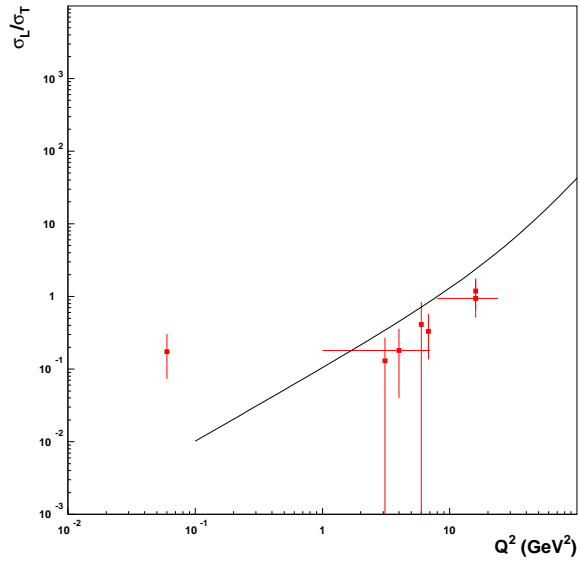}
 \caption{ The ratio $\sigma_L/\sigma_T=R(Q^2)$ as
a function of $Q^2$.
The experimental points are from Refs.~\cite{adloff99,breitweg99,chekanov04}.}
\label{fig3}
\end{figure}

The results are shown in Figs. \ref{fig1}, \ref{fig2} and \ref{fig3}. In  Fig.
\ref{fig1} the differential cross section as a function of $t$ is shown for different
values and $Q^2$ for two values of c.m. energy $W=\sqrt{s}$, $90$ GeV and $110$ GeV.
As one can see, the model correctly describes the data behaviour both in $t$ and in $Q^2$.
This striking agreement shows that the off-mass-shell
generalization of dual amplitude, proposed in Ref. \cite{MDAMA},
works.

The $Q^2$-dependence of $J/\Psi$ electroproduction integrated
elastic  cross section at  $W=90$ GeV is
presented in Fig. \ref{fig2}, left panel. The agreement is rather good.

On the right panel of Fig. \ref{fig2} we study the behaviour of the
integrated elastic cross section of $J/\Psi$ electroproduction as
a function of $W$ for different values of $Q^2$.
The agreement with data is also fairly good, however, one can notice that the fits
deteriorate progressively as $Q^2$ increases.

One reason for this is the violation of the crossing symmetry in the model. Also, the
ratio $R(Q^2)$, Eq. (\ref{Rq}), taken from \cite{martynov}, with
the parameters fitted in that paper, may be too restrictive. The
calculated $R(Q^2)$ is compared to existing experimental data in
Fig. \ref{fig3}.

The agreement is reasonable, however it may be that the parametrization of $R(Q^2)$
of Ref. \cite{martynov}, assumed to be  universal for different vector
mesons: $\rho_0$, $\phi$, $\omega$, $J/\Psi$, should be modified and adjusted to the
particular case of $J/\Psi$.

\section{Conclusion} \label{s4}



In this paper we describe $J/\Psi$ electroproduction in the
framework of M-DAMA Ref.~\cite{jpsi2, MDAMA}. The model has one
free parameter $Q^2_0 $ in the scattering amplitude and the ratio
$\sigma_L/\sigma_T$ is parametrized as in Ref.~\cite{martynov},
with two additional free parameters.
It gives good description of the data in all available regions of
the Mandelstam variables and photon virtualities. The model can be
used to predict cross sections for $J/\Psi$ photo- and
electroproduction for any value of the Mandelstamian variable, and
in a wide range of photon virtualities $Q^2$.

The fits deteriorate at highest values of $Q^2.$ Two reasons for this limitation of the model can be two-fold. One can stem
from the progressive violation of the $s-u$ symmetry, implemented in Eq. (19). This can be cured by adding interference terms,
which however would make the model much more complicated and less practical in applications. The second reason seems obvious:
with increasing $Q^2,$ effects from QCD evolution, ignored in the model, may come into play. We plan to study both effects in
a future paper.

We would like to stress that our results are not restricted to
$J/\Psi\; p$ scattering alone. With a suitable readjustment of
the fitted parameters, the model can be applied to any diffractive
process.

Note that the model produces the smooth universal background in
the near-to-threshold region of the reaction thus it can be used
for the the extraction of tiny resonance signals in the
experiments  with the energies close to the threshold (e.g. at the
JLab).

\section{Acknowledgements}
We thank Prof. F. Paccanoni for fruitful and enlightening
discussions. The work of L.J. was supported by the program
"Fundamental Properties of Physical Systems under Extreme
Conditions" of the Astronomy and Physics Department, National
Academy of Sciences of Ukraine.
V.M. acknowledges the support from
the MICINN (Spain),  contract FIS2008-01661;
from the Ge\-ne\-ra\-li\-tat de Catalunya (Spain), contract 2009SGR-1289;
from CPAN CSD2007-00042 del ProgramaConsolider-Ingenio 2010;
and from the European Community-Research Infrastructure Integrating Activity ``Study of
Strongly Interacting Matter" (acronym HadronPhysics2. Grant
Agreement n. 227431) under the Seventh Framework Programme of EU.
The work was supported in part by the Ministero Italiano
dell'Istruzione, dell'Universit�e della Ricerca.
L.J. thanks the Department of Theoretical Physics of the University of
Calabria and the Istituto Nazionale di Fisica Nucleare - Gruppo Collegato
di Cosenza, where part of this work was done, for their warm hospitality
and support. Authored by Jefferson Science Associates, LLC under U.S. DOE Contract No. DE-AC05-06OR23177. The U.S. Government retains a non-exclusive, paid-up, irrevocable, world-wide license to publish or reproduce this manuscript for U.S. Government purposes.

\end{document}